# High linear low noise amplifier based on self-biasing multiple gated transistors

A.Abbasi, N. Sulaiman
IEEE Member
University Putra Malaysia
a.abbasi.aa@gmail.com

Rozita Teymourzadeh,CEng.
IEEE/IET Member
Neonode Technology
Rozita.Teymourzadeh@neonode.com

*Abstract*— **Noise level frequently set the basic limit on the smallest signal. New noise reduction technology and amplifiers voltage-noise density, yet still offer high speed, high accuracy and low power solution. Low noise amplifiers always play a significant role in RF technology. Hence in this paper, high linear low noise amplifier (LNA) using cascode self-biased multiple gated transistors (MGTR) is presented. The proposed system is covering 0.9 to 2.4 GHz applications. To verify the functionality of the proposed LNA as a bottleneck of RF technology, a cascode LNA without MGTR is implemented and synthesized. The comparison has been done with the single-gate LNA. From the synthesize result, proposed LNA obtained 10 $dBm$ third-order input intercept point (IIP3) in compare with single-gate LNA at 9 $dB$ gain. The proposed LNA is implemented in 90 $nm$ CMOS technology and reported 13 $dBm$ IIP3, 1.9 $dB$ NF and 9 $dB$ gain, while consuming 7.9 $mW$ from 2 V supply.**

*Keywords—linearity, multiple gate transistors linearization, low noise amplifier, LNA, self-biasing.*

## I. Introduction

In the area of RF circuit design, LNAs are critical blocks. Several techniques employed to overcome with nonlinearity and minimized noise figure (NF) in LNAs. By employing differential structure in the input, even-order distortion can be removed easily. However, several techniques used to reduce odd-order nonlinearity [1, 2].

Bruccoleri in 2004 [3] introduced a feed-forward noise-canceling technique to reduce NF. However, the results indicate that linearity and power consumption are not efficient. Kim et.al [2] shows the MGTR method used biasing circuit to remove third-order nonlinearity (second derivative, $g_m''$). However, they are all using extra circuit for biasing and increase complexity of the design. In the same field, Ding [4] proposed a feed-forward linearization technique. Though efficient linearity performance is achieved, the NF frequency band is not reasonable. With circuit technique in the common gate stage based on voltra series analysis [5], IIP3 improvement is achieved. However, for two tone tests, the IIP3 vanishes to 0 dBm.

After peer literature study, proposed research work introduces a self-biased MGTR method to enhance the linearity performance of the LNA. A 90 $nm$ CMOS technology is used to implement self-biased MGTR LNA.

## II. Proposed LNA Circuit Based on Self-Biasing MGTR Method

The common-source (CS) MOSFET is shown in Fig. 1(a). In the CS MOSFET, the nonlinearity comes from $n_{th}$-order derivative of MOSEFT. From Taylor series expansion [2], the drain-source current is expressed as

$$i_{DS} = I_{dc} + g_m v_{gs} + \frac{g_m'}{2!}v_{gs}^2 + \frac{g_m''}{3!}v_{gs}^3 + \cdots \quad (1)$$

where $g_m'$ and $g_m''$ are the first and second-order derivative, respectively. It is clear that $v_{gs}^3$ play critical role in the third-order distortion of RF circuits. The $g_m'$ and $g_m''$ of the simple CS NMOSFET shown in Fig. 1(b). It indicates that $g_m''$ goes to negative peak value at voltage larger than $V_{th}$, degrading the linearity of the amplifier. In MGTR method, this negative peak value of main transistor (MT) can be removed by another transistor called second transistor (ST). To achieve this propose, different biasing voltage should be applied to the gate of MT and ST [1, 2].

### A. IIP3 of the proposed LNA

In this paper, a self-biasing circuit without external biasing circuit is proposed in Fig. 2(a) to reduce complexity of the design and increase linearity. In proposed circuit, MT is the main transistor and ST is the second or auxiliary transistor. PMOS is chosen for input transistor due to less sensitivity of PMOS to the noise. Control transistor (CT) is the NMOS current-control transistors to control the $V_{gs}$ of the main and second transistors and act as resistor. Firstly, the bulk of the CT transistor, which is connected to the MT and ST transistor, is connected to the source. However, to set the point for MT and ST transistors to remove $g_m''$, W/L ratio of CT for ST should be significant and W/L ratio of CT for MT should be tiny. To address the problem, if the balk of CT for ST connect to the $V_{SS}$, $V_{th} = V_{TO} + \gamma(\sqrt{V_{SB} + 2\varphi_B} - \sqrt{2\varphi_B})$ increases and $(V_{gs} - V_{th})$ decreases. Therefore, due to equ. (7) higher value of resistor achieved for ST. Hence, reasonable W/L ratio for CT transistor can set point for the $g_m''$ of MT and ST to make MT+ST close to zero value.

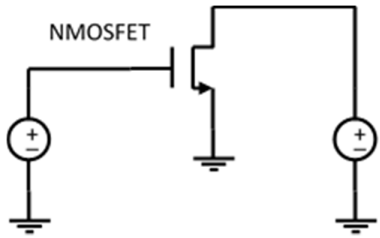
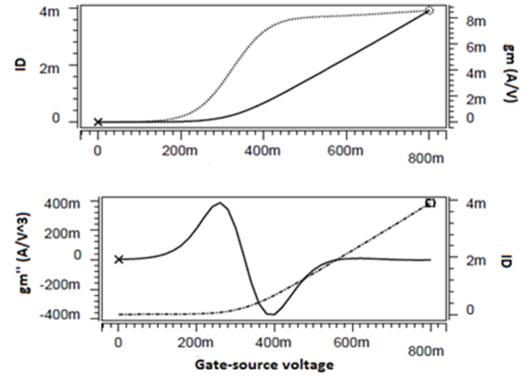

a) Common-source (CS) MOSFET.

b) $g'_m$ and $g''_m$ of CS MOSFET.

Fig. 1. Common-source MOSFET with first and second derivative characteristics.

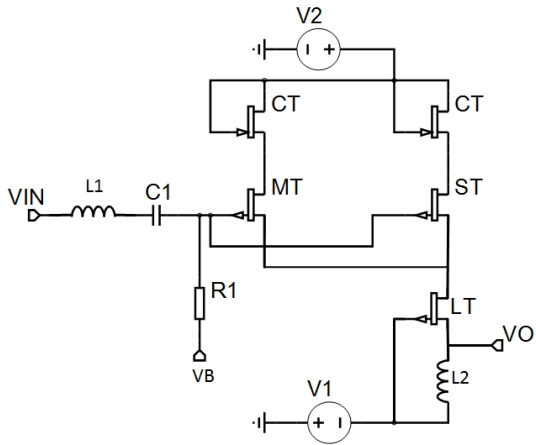
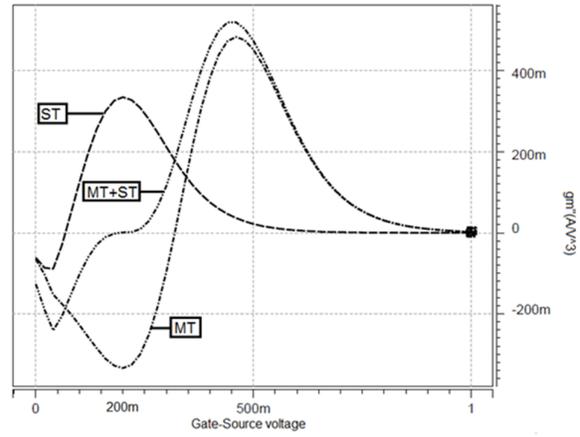

a) Proposed self-biased MGTR LNA.

b) $g''_m$ of ST, MT and their superposing.

Fig. 2. Proposed LNA with first and second derivative characteristics.

By modifying W/L ratio of the CT, the transfer function characteristics of the MT and ST is shifted to find the exact bias point to remove $g''_m$ of each other. As seen in Fig. 2(b), $g''_m$ of the MT+ST is vanished to zero value. LT as a cascode transistor is used to increase IIP3 based on equ. (5) [2]. In the cascode configuration, $Z_2$ is decreased to $\frac{1}{g_m}$ and causes IIP3 improvement. $L_2$ is an inductive load and causes operating LNA at higher frequency. Moreover, as shown in equ. (6), the gain also related to $L_2$. Thus, $L_2$ should be as large as possible to achieve higher gain. The $L_1$ is the matching network that operates LNA in different frequency. The $V_B$ is the biasing voltage for both transistors. Extra biasing voltage is not required for each transistors due to the $V_{gs}$ can be modified by CT.

$$IIP_3(2\omega_a - \omega_b) = \frac{1}{6Re[Z_s(\omega)]|H(\omega)||A_1(\omega)|^3|\varepsilon(\Delta\omega, 2\omega)|} \quad (2)$$

$$\varepsilon(\Delta\omega, 2\omega) = g_3 - g_{OB} \quad (3)$$

$$g_{OB} = \frac{2g_2^2}{3}\left[\frac{2}{g_1 + g(\Delta\omega)} + \frac{1}{g_1 + g(2\omega)}\right] \quad (4)$$

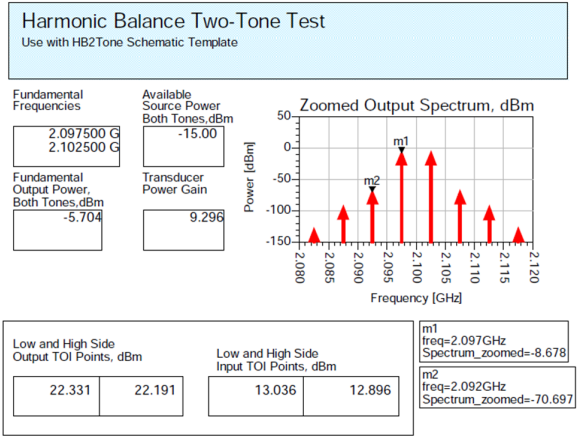 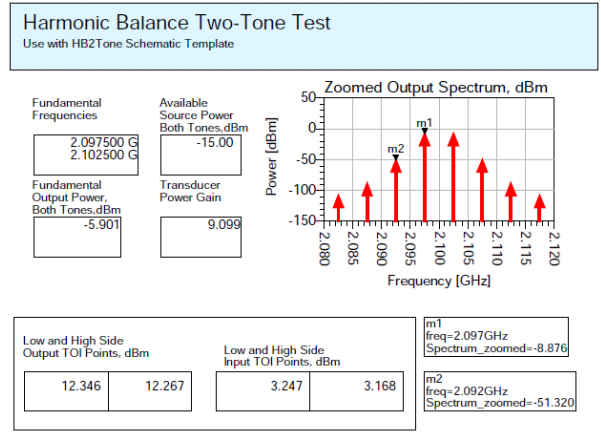

a) IIP3, IMD3 and gain of the proposed LNA at 2.1 GHz.  b) IIP3, IMD3 and gain of the single-gate LNA at 2.1 GHz.

Fig. 3. The comparison between the proposed LNA and Single-gate LNA at 2.1 GHz.

$$g(2\omega) \approx g_m \times \frac{1 + 2j\omega C_{gs}Z_1 + 2j\omega C_{gd}Z_2}{1 + \omega_T C_{gd}Z_2} \quad (5)$$

$$A_{v,LNA} = g_m Z_{out} \quad (6)$$

$$R_{AB} = \frac{L}{KW(V_{gs} - V_{th} - V_{ds})} \quad (7)$$

The IIP3 is not only related to the $g_m''$, also $g_m'$ degrade IIP3. In equ. (5), the amplitude of $Z_1$ is of the order of $(1/\omega C_{gs})$. Therefore, it does not have impressive influence. The second term in the numerator of (5) is comparable to 1. Moreover, it is well known [6] that $2\omega \ll \omega_T$, hence, the term of $\omega_T C_{gd}Z_2$ In the denominator of (5) is the most effective factor. Therefore, $\omega_T C_{gd}Z_2$ should be reduced smaller than 1. Thus, in (5) $Z_2$ should be decreased dramatically. In proposed research work, cascode configuration is employed to reduce $Z_2$. $Z_2$ is decreased to $1/g_m$ in cascode configuration. Thus, the W/L ratio of LT should be large enough to improve the IIP3.

*B. Noise Calculation*

It is well known [7] that the noise current can be modeled by voltage source in series with gate. There are two types of noise that affect the MOSFET, flicker noise and thermal noise [7]. The effect of flicker noise in high frequency is negligible. Thus, thermal noise is investigated [7, 8].
The current noise and voltage noise for MOSFET are given by [7].

$$\overline{I_n^2} = 4kT\gamma g_m, \quad \overline{v_{n,in}^2} = 4kT\gamma/g_m \quad (8)$$

Where $\gamma$ is the ''excess noise coefficient'' and $g_m$ is the transconductance. The value of $\gamma$ is 2/3 for long channel transistors, K is Boltzmann's constant, T is the absolute temperature.
The NF for the LNA is calculated by [7].

$$NF = 1 + \frac{\overline{v_{n,in}^2}}{4kTR_S} \quad (9)$$

Because MT and ST are in parallel,

$$NF = 1 + 4kT\gamma \left( \frac{\left(\frac{1}{g_{m,MT(eff)}} + \frac{g_{m,LT}}{g_{m,MT(eff)}^2}\right)}{4kTR_S} + \frac{\left(\frac{1}{g_{m,ST(eff)}} + \frac{g_{m,LT}}{g_{m,ST(eff)}^2}\right)}{4kTR_S} \right) \quad (10)$$

where $g_{m,MT(eff)} = \frac{g_{m,MT}}{1 + g_{m,MT}/g_{m,CT}}$, $g_{m,ST(eff)} = \frac{g_{m,ST}}{1 + g_{m,ST}/g_{m,CT}}$ and $R_S$ is source load equal to 50Ω [9]. Furthermore, $g_{m,ST} \ll g_{m,MT}$, thus, $4kT\gamma/g_{m,ST}$ is not dominating factor.

$$NF \approx 1 + \frac{\gamma}{R_S g_{m,MT(eff)}} \quad (11)$$

III. SIMULATION RESULTS

The proposed LNA is implemented using 90 $nm$ CMOS technology for 0.9-2.4 GHz applications. In Fig. 3(a) and Fig. 3(b), the comparison of third-order intermodulation distortion (IMD3), IIP3 and gain of the LNA at 2.1 GHz between single-gate cascode LNA and proposed LNA show 10 dB IIP3 improvement after using MGTR method with the same gain. By applying two-tone test at 2.1 GHz center frequency with 5 MHz variety and -15 dBm input power, (Fig. 3) is achieved. IIP3 and gain are 13 dBm and 9 dB, respectively. 1.9 dB NF is shown in Fig. 4 at 2.1 GHz. Table 1. shows the comparison of the measurement results with some existing research works.

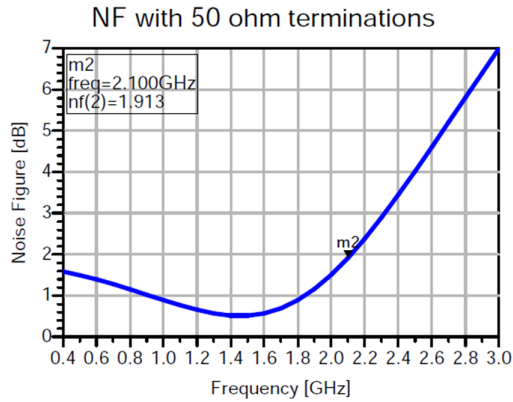

Fig. 4. NF of the proposed LNA at 2.1 GHz.

Table 1. Comparison of the Proposed LNA with Previous Works

|  | This Work | [10] | [11] | [12] | [13] |
|---|---|---|---|---|---|
| Technology | 90 nm CMOS | 0.18 μm CMOS | 0.13 μm CMOS | 45 nm CMOS | 90 nm CMOS |
| Supply (V) | 2 | 1.8 | 1 | 2.2 | 2.5 |
| Freq Offset [GHz] | 2.1 | 3.1 | 4 | 1 | 0.002-1.1 |
| BW [GHz] | 0.9-2.4 | 3.1-10.6 | 4 | 0.1-2 | 0.002-1.1 |
| Gain [dB] | 9 | 20 | 11.2 | -1.7 | 20 |
| IIP3 [dBm] | 13 | 10 | 1.5 | 10.8 | -1.5 |
| NF | 1.9 | 2.89 | 2.4-3 | 3 | 1.43 |
| Single (S) or Differential (D) | S | S | D | D | D |
| Pdc [mW] | 7.9 | 12 | 16.9 | 30.2 | 18 |

## IV. CONCLUSION

An LNA using self-biased MGTR method is proposed in this paper. The LNA relies on a self-biased MGTR for cancelling third-order harmonic distortion and improve linearity. To verify the proposed LNA linearity performance, a cascode LNA without MGTR is designed and implemented for test and analyzing. The comparison has been made to prove the proposed LNA efficiency. The comparison results illustrate the new algorithm achieved 10 $dBm$ IIP3 improvement in compare with single-gate LNA at 9 $dB$ gain. The proposed LNA is implemented in 90 $nm$ CMOS technology to realize 13 $dBm$ IIP3, 1.9 $dB$ NF and 9 $dB$ gain, while consuming 7.9 $mW$ from 2 V supply.